# Validation of a Third Planet in the LHS 1678 System

Michele L. Silverstein[1,2,3,22], Thomas Barclay[2], Joshua E. Schlieder[2], Karen A. Collins[4], Richard P. Schwarz[4], Benjamin J. Hord[2,23], Jason F. Rowe[5], Ethan Kruse[2,6], Nicola Astudillo-Defru[7], Xavier Bonfils[8], Douglas A. Caldwell[9,10], David Charbonneau[4], Ryan Cloutier[11], Kevin I. Collins[12], Tansu Daylan[13], William Fong[14], Jon M. Jenkins[15], Michelle Kunimoto[14], Scott McDermott[16], Felipe Murgas[17,18], Enric Palle[17,18], George R. Ricker[14], Sara Seager[14,19,20], Avi Shporer[14], Evan Tey[14], Roland Vanderspek[14], and Joshua N. Winn[21]

[1] University of Maryland, Baltimore County, 1000 Hilltop Circle, Baltimore, MD 21250, USA; michele.l.silverstein.ctr@us.navy.mil  
[2] NASA Goddard Space Flight Center, Greenbelt, MD 20771, USA  
[3] GSFC Sellers Exoplanet Environments Collaboration, Greenbelt, MD 20771, USA  
[4] Center for Astrophysics | Harvard & Smithsonian, 60 Garden Street, Cambridge, MA, 02138, USA  
[5] Department of Physics and Astronomy, Bishops University, 2600 Rue College, Sherbrooke, QC J1M 1Z7, Canada  
[6] University of Maryland, College Park, MD 20742, USA  
[7] Departamento de Matemática y Física Aplicadas, Universidad Católica de la Santísima Concepción, Alonso de Rivera 2850, Concepción, Chile  
[8] Univ. Grenoble Alpes, CNRS, IPAG, F-38000 Grenoble, France  
[9] SETI Institute, Mountain View, CA 94043 USA  
[10] NASA Ames Research Center, Moffett Field, CA 94035 USA  
[11] Department of Physics & Astronomy, McMaster University, 1280 Main Street West, Hamilton, ON, L8S 4L8, Canada  
[12] George Mason University, 4400 University Drive, Fairfax, VA, 22030 USA  
[13] Department of Physics and McDonnell Center for the Space Sciences, Washington University, St. Louis, MO 63130, USA  
[14] Department of Physics and Kavli Institute for Astrophysics and Space Research, Massachusetts Institute of Technology, Cambridge, MA 02139, USA  
[15] NASA Ames Research Center, Moffett Field, CA 94035, USA  
[16] Proto-Logic LLC, 1718 Euclid Street NW, Washington, DC 20009, USA  
[17] Instituto de Astrofísica de Canarias (IAC), E-38205 La Laguna, Tenerife, Spain  
[18] Departamento de Astrofísica, Universidad de La Laguna (ULL), E-38206 La Laguna, Tenerife, Spain  
[19] Department of Earth, Atmospheric, and Planetary Sciences, Massachusetts Institute of Technology, Cambridge, MA 02139, USA  
[20] Department of Aeronautics and Astronautics, MIT, 77 Massachusetts Avenue, Cambridge, MA 02139, USA  
[21] Department of Astrophysical Sciences, Princeton University, 4 Ivy Lane, Princeton, NJ 08544, USA  
Received 2023 July 31; revised 2024 February 15; accepted 2024 February 20; published 2024 May 6

## Abstract

The nearby LHS 1678 (TOI-696) system contains two confirmed planets and a wide-orbit, likely brown-dwarf companion, which orbit an M2 dwarf with a unique evolutionary history. The host star occupies a narrow "gap" in the Hertzsprung–Russell diagram lower main sequence, associated with the M dwarf fully convective boundary and long-term luminosity fluctuations. This system is one of only about a dozen M dwarf multiplanet systems to date that hosts an ultra-short-period planet (USP). Here we validate and characterize a third planet in the LHS 1678 system using TESS Cycle 1 and 3 data and a new ensemble of ground-based light curves. LHS 1678 d is a $0.98 \pm 0.07 \, R_\oplus$ planet in a 4.97 day orbit, with an insolation flux of $9.1^{+0.9}_{-0.8} \, S_\oplus$. These properties place it near 4:3 mean motion resonance with LHS 1678 c and in company with LHS 1678 c in the Venus zone. LHS 1678 c and d are also twins in size and predicted mass, making them a powerful duo for comparative exoplanet studies. LHS 1678 d joins its siblings as another compelling candidate for atmospheric measurements with the JWST and mass measurements using high-precision radial velocity techniques. Additionally, USP LHS 1678 b breaks the "peas-in-a-pod" trend in this system although additional planets could fill in the "pod" beyond its orbit. LHS 1678's unique combination of system properties and their relative rarity among the ubiquity of compact multiplanet systems around M dwarfs makes the system a valuable benchmark for testing theories of planet formation and evolution.

*Unified Astronomy Thesaurus concepts:* Exoplanet systems (484); Transit photometry (1709); Transit timing variation method (1710); M dwarf stars (982)

*Supporting material:* machine-readable table

## 1. Introduction

As the Transiting Exoplanet Survey Satellite (TESS; Ricker et al. 2015) mission continues into its sixth year, discovering over 350 confirmed and 6500 candidate planet candidates to date,[24] the exoplanet science community is steadily accumulating a diverse ensemble of systems comprised of small planets orbiting bright, nearby M dwarfs (with ensemble studies and TESS discoveries by, e.g., Gilbert et al. 2020, 2023; Giacalone et al. 2022; Luque & Pallé 2022; Ment & Charbonneau 2023; Rodríguez Martínez et al. 2023; and many more). These systems are key for detailed follow-up using cutting-edge observatories such as the James Webb Space Telescope (JWST; Gardner et al. 2006) to begin probing and piecing together the puzzle of exoplanet atmospheres and searching for planets like the Earth. One of these bright, nearby systems ripe for space-based follow-up is the LHS 1678 exoplanet system, first discovered and characterized by Silverstein et al. (2022). LHS 1678 is one of few exoplanet host stars associated with the M-dwarf fully convective boundary, identified as an underdensity or "gap" in the Hertzsprung–Russell (H-R) diagram (van Saders & Pinsonneault 2012; Jao et al. 2018; Feiden et al. 2021). Gap stars are predicted to exhibit characteristic luminosity oscillations with periods of ∼200 Myr to ∼2 Gyr that persist for billions of years (Feiden et al. 2021); implications for exoplanet demographics, climate, and habitability remain under investigation (M. L. Silverstein et al. 2024b, in preparation; I. Urquhart et al. 2024, in preparation). The host star is orbited by a likely brown dwarf, two confirmed small planets, and a third planet candidate. The properties of the substellar companion, first identified by Jao et al. (2017), remain uncertain; we await additional astrometric observations to append to the 16 yr of data presented by Silverstein et al. (2022) to unveil the orbital properties and provide improved limits on the mass of the companion.

---

[22] Now resident at the U.S. Naval Research Laboratory as an NRC Research Associate; michele.silverstein.ctr@nrl.navy.mil.  
[23] NASA Postdoctoral Program Fellow.

[24] NASA Exoplanet Archive: https://exoplanetarchive.ipac.caltech.edu.

Silverstein et al. (2022) identified the remaining candidate as a faint transit signal in TESS Cycle 1 and 3 data and identified the need for ground-based light curves to validate it. They showed that the planet candidate and LHS 1678 c have periods within 1% of a 4:3 mean motion resonance. Systems close to resonance allow for the detection of transit timing variations (TTVs; Agol et al. 2005; Holman & Murray 2005), which can provide access to further properties of the planets like eccentricity and mass. TTV characterization is also important for pinning down system ephemerides used to plan future observations and can reveal the influence of additional planets in the system (discussed by, e.g., Ford et al. 2011; Baştürk et al. 2022, and references therein).

Here we present new TESS observations and additional ground-based follow-up of the LHS 1678 planets. We focus our sights on confirming the candidate TOI-696.03 and refine the properties of all three planets with the available data. We also search for TTVs using the TESS Cycle 1 and 3 data sets and the complement of ground-based transits. With the validation of planet d and refined system properties, we comment on the interesting properties of this new planet in the context of the system and its prospects for future characterization.

## 2. Observations

We employ the full TESS Cycle 1 and 3 data sets, a suite of 33 ground-based light curves from the Las Cumbres Observatory global telescope (LCOGT; Brown et al. 2013) network and MEarth-South telescope array (Nutzman & Charbonneau 2008; Berta et al. 2012; Irwin et al. 2015) and radial velocities from the High Accuracy Radial Velocity Planet Searcher (HARPS) echelle spectrograph (Pepe et al. 2002; Mayor et al. 2003).

### 2.1. TESS

LHS 1678 (TIC 77156829, TOI-696, L 375-2, LTT 2022, NLTT 13515) was observed at 2 minute cadence by TESS in Cycle 1 Sectors 4 and 5 using Camera 3 from UT 2018 October 19 to 2018 November 14 and in Cycle 3 Sectors 31 and 32 using Camera 3 from UT 2020 October 21 to 2020 December 17 as part of Cycle 1 Guest Investigator Program G011180,[25] the Cool Dwarf target catalog (Muirhead et al. 2018),[26] and the TESS Candidate Target List (CTL; Stassun et al. 2018) and Cycle 3 Guest Investigator Programs G03228,[27] G03272,[28] G03274,[29] and G03278,[30] the Cool Dwarf target catalog, and the TESS CTL.

The TESS photometry was processed into light curves by the NASA Ames Science Processing Operations Center (SPOC; Jenkins et al. 2016) pipeline. No evidence of starspot modulation or flares was evident in the 4 months of data. Two planet signals were identified in the Cycle 1 data using the Transiting Planet Search (TPS; Jenkins et al. 2010) module, which were confirmed and validated by Silverstein et al. (2022) as LHS 1678 b and c. These planets were also distinguishable in the TESS Cycle 3 data, and with the combination of Cycle 1 and 3 data, a third planet was independently identified in two full-frame image searches using a Box-fitting Least Squares (Kovács et al. 2002; Vanderburg et al. 2016) algorithm, the Quick-Look Pipeline (QLP; Huang et al. 2020a, 2020b), and the TPS module.

### 2.2. Ground-based Time-series Observations

The LHS 1678 planets were observed over 3 yr using multiple telescopes. We focused our efforts on ground-based photometric time-series observations with two goals: to measure multiple transits of each planet in the system and to search for TTV signatures over a significant time baseline. Most of these observations were gathered using telescopes in the LCOGT network.

Using the 1.0 m telescopes at three LCOGT sites in the Southern Hemisphere, we acquired ten transit observations of LHS 1678 b, thirteen transit observations of LHS 1678 c, and nine transit observations of the candidate planet TOI-696.03. Each telescope hosts a Sinistro CCD camera with a pixel scale of 0″.39 pixel$^{-1}$ and 26′ field of view. Observations were performed in the $zs$, $I_c$, or $ip$ filter. We calibrated the images using the LCOGT `Banzai` pipeline (McCully et al. 2018). The `AstroImageJ` software package (Collins et al. 2017) was used to extract differential photometry from the images and generate light curves. For TOI-696.03, we also searched for and ruled out eclipsing binaries (EBs) at the candidate orbital period. All LCOGT light curves were newly analyzed since Silverstein et al. (2022), using `AstroImageJ v5`,[31] the latest version and most optimized to date for transiting planet light-curve construction. The resulting LCOGT light curves are incorporated in the modeling of all three exoplanets to derive parameters and the search for TTVs, both described in Section 3.1.

Additional observations, described in Silverstein et al. (2022), include one LHS 1678 c transit observed using the MEarth-South telescope system. Table 1 lists the full suite of ground-based LCOGT and MEarth light curves incorporated in our exoplanet modeling procedure.

---

[25] Differential Planet Occurrence Rates for Cool Dwarfs (PI: C. Dressing). Details of approved TESS Guest Investigator Programs are available from https://heasarc.gsfc.nasa.gov/docs/tess/approved-programs.html.
[26] http://vizier.u-strasbg.fr/viz-bin/VizieR?-source=J/AJ/155/180
[27] High Frequency Quasi-Periodic Pulsations In M Dwarf Flares (PI: C. Million)
[28] Two Minute TESS Data For Thousands Of Promising New Radial Velocity Target Stars (PI: J. Burt)
[29] Understanding The Physical Origin Of The Rocky/Non-Rocky Transition Around Mid-To-Late M Dwarfs With TESS (PI: R. Cloutier)
[30] Enriching Our View Of Multiplanet Systems Using TESS (PI: A. Mayo)
[31] http://astroimagej.170.s1.nabble.com/AstroImageJ-version-5-installation-and-overview-of-new-feature-operation-td1729.html

Table 1
Ground-based Time-series Observations[a]

| Telescope | Epoch (UT) | Filter | Exposure (s) | $R_{aperture}$ (pixel) | Duration (minute) | # of Obs. | Reference[b] |
|---|---|---|---|---|---|---|---|
| LHS 1678 b | | | | | | | |
| LCO CTIO 1.0 m | 2019-07-27 | zs | 20 | 15 | 125 | 163 | Sil22 |
| LCO SSO 1.0 m | 2019-08-06 | zs | 40 | 15 | 110 | 100 | This work |
| LCO CTIO 1.0 m | 2019-12-05 | Ic | 30 | 15 | 185 | 197 | Sil22 |
| LCO SSO 1.0 m | 2019-12-14 | zs | 60 | 15 | 190 | 129 | This work |
| LCO SAAO 1.0 m | 2019-12-24 | zs | 60 | 15 | 192 | 128 | Sil22 |
| LCO SAAO 1.0 m | 2020-07-26 | zs | 60 | 20 | 132 | 86 | Sil22 |
| LCO SAAO 1.0 m | 2020-08-20 | zs | 60 | 16 | 204 | 133 | Sil22 |
| LCO SSO 1.0 m | 2020-09-28 | zs | 60 | 20 | 226 | 146 | Sil22 |
| LCO CTIO 1.0 m | 2020-11-25 | ip | 23 | 20 | 216 | 225 | This work |
| LCO SSO 1.0 m | 2020-11-30 | ip | 23 | 25 | 172 | 181 | This work |
| LHS 1678 c | | | | | | | |
| LCO SSO 1.0 m | 2019-08-14 | zs | 40 | 15 | 178 | 161 | This work |
| LCO CTIO 1.0 m | 2020-02-08 | zs | 40 | 15 | 243 | 195 | This work |
| MEarth-South-x7-0.4 m | 2020-03-16 | RG715 | 60 | 9.9 | 239 | 1170 | Sil22 |
| LCO SSO 1.0 m | 2020-11-18 | zs | 60 | 20 | 339 | 215 | This work |
| LCO SAAO 1.0 m | 2020-11-25 | zs | 60 | 15 | 291 | 188 | This work |
| LCO SAAO 1.0 m | 2021-02-07 | ip | 25 | 20 | 236 | 239 | This work |
| LCO CTIO 1.0 m | 2021-10-02 | ip | 23 | 20 | 195 | 219 | This work |
| LCO CTIO 1.0 m | 2021-11-08 | ip | 23 | 15 | 196 | 225 | This work |
| LCO SAAO 1.0 m | 2021-11-11 | ip | 23 | 15 | 173 | 193 | This work |
| LCO SSO 1.0 m | 2021-11-15 | ip | 23 | 20 | 195 | 214 | This work |
| LCO CTIO 1.0 m | 2021-12-04 | ip | 23 | 15 | 195 | 220 | This work |
| LCO SSO 1.0 m | 2021-12-11 | ip | 23 | 15 | 152 | 170 | This work |
| LCO CTIO 1.0 m | 2021-12-15 | ip | 23 | 15 | 196 | 224 | This work |
| LCO SAAO 1.0 m | 2021-12-29 | ip | 23 | 15 | 196 | 225 | This work |
| LHS 1678 d | | | | | | | |
| LCO SSO 1.0 m | 2021-09-09 | ip | 23 | 15 | 194 | 222 | This work |
| LCO CTIO 1.0 m | 2021-11-03 | ip | 23 | 15 | 184 | 212 | This work |
| LCO CTIO 1.0 m | 2021-11-18 | ip | 23 | 15 | 184 | 210 | This work |
| LCO CTIO 1.0 m | 2021-11-23 | ip | 23 | 15 | 184 | 181 | This work |
| LCO CTIO 1.0 m | 2021-11-28 | ip | 23 | 15 | 185 | 210 | This work |
| LCO CTIO 1.0 m | 2021-12-03 | ip | 23 | 15 | 184 | 141 | This work |
| LCO SAAO 1.0 m | 2021-12-17 | ip | 23 | 20 | 184 | 209 | This work |
| LCO SAAO 1.0 m | 2022-01-01 | ip | 23 | 15 | 185 | 214 | This work |
| LCO SAAO 1.0 m | 2022-01-06 | ip | 23 | 15 | 183 | 209 | This work |

**Notes.**
[a] All ground-based transit observations were observed as continuous time series. LCO and MEarth observations were taken with pixel scales of $0.''39$ pixel$^{-1}$ and $0.''84$ pixel$^{-1}$, respectively. Additional details about the observations are available on ExoFOP (https://exofop.ipac.caltech.edu). Observations on 2019 July 27, 2019 December 14, 2021 November 8, 2021 November 15, 2021 November 18, and 2022 January 1 capture full or partial transits of multiple planets (b&c, b&c, c&d, b&c, b&d, and b&d, respectively) and are listed here under the intended target planet.
[b] New observations presented for the first time in this publication are labeled "This work." Observations used in Silverstein et al. (2022) are labeled "Sil22."

Silverstein et al. (2022) also detail radial velocity time series acquired using the HARPS instrument. These data were acquired via ESO Programme ID: 1102.C-0339 and used to constrain the masses of planets b and c. Here, we use the same data in a joint analysis including the candidate third planet (Section 3.1).

## 3. Data Analysis and Results

Here we (1) update the properties of all three planets with the full suite of TESS data, ground-based light curves, and HARPS radial velocities; (2) perform a search for TTVs; and (3) validate TOI-696.03.

We adopt stellar properties from Silverstein et al. (2022), with key values listed in Table 2. Updated exoplanet properties are reported in Table 3 and described in Section 3.1.

### 3.1. Joint Modeling of Space and Ground-based Data

We built a probabilistic model of the orbit of the three planets around LHS 1678 using the time-series photometry from TESS and the ground, along with the radial velocity data from HARPS. This model was quite similar to that used in Silverstein et al. (2022). We begin with the TESS Presearch Data Conditioned Simple Aperture Photometry (PDCSAP; Smith et al. 2012; Stumpe et al. 2012, 2014) light curves extracted and processed by the NASA Ames SPOC pipeline (Jenkins et al. 2016).

We estimate the planet properties using the procedure described by Silverstein et al. (2022).

We use the exoplanet package (Foreman-Mackey et al. 2020) to approximate the exoplanet properties, jointly modeling TESS and ground-based light-curve transits, stellar variability, and other systematics, in addition to the HARPS RV data. Simultaneous modeling of all data sets maintain covariances between parameters. Table 1 lists the ground-based photometry included in the model, and the HARPS RVs are reported by Silverstein et al. (2022).

**Table 2**
LHS 1678 Properties from Silverstein et al. (2022)

| Property | Value | Error |
|---|---|---|
| Mass ($M_\odot$) | 0.345 | 0.014 |
| Rotation Period (days) | 64 | 22 |
| Effective Temperature (K) | 3490 | 50 |
| Bolometric Flux (log erg cm$^{-2}$ s$^{-1}$) | −8.932 | 0.008 |
| Luminosity ($L_\odot$) | 0.0145 | 0.0003 |
| Radius ($R_\odot$) | 0.329 | 0.010 |
| Density (g cm$^{-3}$) | 13.624 | 1.40 |
| Age (Gyr) | 4–9 | ... |

Each TESS sector and ground-based light curve is modeled with a mean offset and additional term to capture unreported observational uncertainty. Remaining stellar variability is modeled using a damped simple harmonic oscillator Gaussian process (GP) model (Foreman-Mackey et al. 2017; Foreman-Mackey 2018). The TESS data and ground-based light curves are modeled using the same GP, with independent hyperparameters and an additional parameter to capture uncertainty between different observatories. For the RV model, we used a log-normal prior on radial velocity semiamplitude with a mean of log(1 m s$^{-1}$) and standard deviation of 5 dex. We also employed a jitter term and quadratic trend. Stellar property inputs are identical to Silverstein et al. (2022) with the exception of TESS and Las Cumbres Observatory (LCO) limb-darkening parameters, which change with the additional data. In this model, we also opt to include the central transit time ($T_c$) rather than the midpoint time of first transit (Deeg 2015). $T_c$ is the midpoint of the transit in the middle of the timespan of the observation period. There is negligible difference in our results. $T_c$ provides a more recent ephemeris for predicting future transits. Prior probability distributions for each modeled parameter are the same as those in Silverstein et al. (2022).

We carried out Hamiltonian Monte Carlo sampling from the posterior model using PyMC3 (Salvatier et al. 2016), as in Silverstein et al. (2022). We phase-fold the central 68th percentile of transit models for the TESS data in Figure 1. Ground-based transit light curves and the associated model fits for planets b, c, and d are similarly displayed in Figures 2, 3, 4, and 5.

The incorporation of additional TESS Sectors (31 and 32) and significant additional ground-based observations have led to better constraints on the parameters of planets b and c and the first full constraints on planet d. We also attempted to place constraints on the masses of the three planets. However, with no additional radial velocity data obtained since the initial system discovery paper and the added degeneracy introduced by the third planet, our model converges on masses consistent with zero.

We provide in Table 3 the median model parameters and 1$\sigma$ uncertainties computed during the PyMC3 sampling. The properties of LHS 1678 b and c have changed minimally from Silverstein et al. (2022). However, we recommend that the community use the updated values in this work when considering future follow-up or related research programs because of the larger number of observed transits spanning a longer time baseline. The $T_c$, the midpoint time of the middle transit, is also a more recent time for the ephemerides.

### 3.2. Modeling Individual Transit Times

Silverstein et al. (2022) predicted that LHS 1678 c and the planet candidate could exhibit TTVs with a superperiod of ∼155 days and amplitudes of ∼2 minutes or larger. To place constraints on potential TTVs in the system, we also built a model that did not force the orbits of the planets to have a strictly linear ephemeris. This was done using the TTVOrbit method within the exoplanet software (Foreman-Mackey et al. 2021). TTVOrbit works by continuing to use a Keplerian orbit but warping the time vector to allow the transit times to vary. To speed up the model, and because we were primarily only interested in measuring the transit times, we simplified it to set eccentricity to zero and fixed the stellar parameters and transit depths. The transit times used Gaussian priors. The prior mean for the transit times was calculated from the linear ephemeris model, and standard deviation was assumed to be 1 hr. We do not see any evidence for TTVs (Figure 6). Because these are small planets around a relatively faint M dwarf, the signal-to-noise of individual transits is low. The individual transit times are therefore measured with average uncertainties of order 0.01 days or 14.4 minutes (Table 4). The ratio between O-C value and transit time uncertainty reaches a maximum of only 1.99, 2.37, and 2.38 for LHS 1678 b, c, and d, respectively, less than three for all transits for all planets. The scatter on the TTVs (standard deviations of 0.004, 0.006, and 0.020 days or 5.76, 8.64, and 28.8 minutes for b, c, and d, respectively) is also consistent with the measurement uncertainty on the transit times (means of 0.007, 0.009, and 0.020 days or 10.1, 13.0, and 28.8 minutes). Therefore, we do not detect TTV in our data. Significantly better precision on individual transit measurements using ground- or space-based instruments is needed to further constrain TTVs in the system.

### 3.3. Statistical Validation

In Silverstein et al. (2022) we used the host star properties, constraints on the presence of EBs, and wider-orbit companions, and the TESS transit data to calculate false positive probabilities (FPPs) and statistically validate LHS 1678 b and c using vespa. For continuity with that work, here we perform the same analysis, with the same input constraints, the revised system properties from our joint modeling, and the full set of TESS data, on the candidate TOI-696.03. The analysis returns an FPP of $<10^{-6}$. When combined with the multiplanet system multiplicity boost from Guerrero et al. (2021), the FPP is vanishingly small.

Recently, Morton et al. (2023) recommended transitioning away from using vespa in favor of TRICERATOPS, citing that vespa is no longer being maintained and updated. Thus, we also performed an independent validation of the signal through the statistical validation software TRICERATOPS (Giacalone & Dressing 2020; Giacalone et al. 2021). This software is similar to vespa in that it compares the user-provided phase-folded light curve, stellar parameters, and planetary parameters against a set of astrophysical false positive scenarios to rule out portions of parameters space in which each scenario is viable. Since TRICERATOPS was developed initially for use with TESS data, it accounts for the real sky background out to 2.′5, extraction aperture, and TESS point-spread function, which vespa does not.

Table 3
Planet Parameters

| Parameter | Median | $+1\sigma$ | $-1\sigma$ |
|---|---|---|---|
| **Measured Parameters** | | | |
| **Star** | | | |
| $\rho$ (g cm$^{-3}$) | 13.5 | 1.5 | 1.4 |
| Limb-darkening TESS $u_1$ | 0.98 | 0.39 | 0.43 |
| Limb-darkening TESS $u_2$ | −0.18 | 0.46 | 0.36 |
| Limb-darkening LCO $u_1$ | 0.75 | 0.44 | 0.40 |
| Limb-darkening LCO $u_2$ | −0.03 | 0.36 | 0.31 |
| Limb-darkening MEarth $u_1$ | 0.95 | 0.56 | 0.61 |
| Limb-darkening MEarth $u_2$ | −0.21 | 0.54 | 0.43 |
| **LHS 1678 b** | | | |
| Central transit ($T_c$) (BJD-2457000) | 1998.15553 | 0.00070 | 0.00071 |
| Period (day) | 0.8602325 | 0.0000013 | 0.0000011 |
| Impact parameter† | 0.21 | 0.16 | 0.14 |
| $R_p/R_*$ | 0.01905 | 0.00087 | 0.00080 |
| Radius ($R_\oplus$) | 0.685 | 0.037 | 0.035 |
| K (m s$^{-1}$) | 0.92 | 0.72 | 0.91 |
| eccentricity† | 0.033 | 0.035 | 0.023 |
| $\omega$ (radian)† | −0.4 | 2.5 | 1.8 |
| $a/R_*$ | 8.08 | 0.29 | 0.28 |
| $a$ (au) | 0.01239 | 0.00056 | 0.00057 |
| Inclination (deg) | 88.53 | 1.02 | 1.14 |
| Duration (hr) | 0.801 | 0.046 | 0.049 |
| Insolation ($S_\oplus$) | 94.3 | 9.1 | 8.1 |
| **LHS 1678 c** | | | |
| Central transit ($T_c$) (BJD-2457000) | 1998.45607 | 0.00064 | 0.00057 |
| Period (days) | 3.6942840 | 0.0000048 | 0.0000044 |
| Impact parameter† | 0.44 | 0.09 | 0.14 |
| $R_p/R_*$ | 0.0262 | 0.0011 | 0.0012 |
| Radius ($R_\oplus$) | 0.941 | 0.051 | 0.050 |
| K (m s$^{-1}$) | 0.01 | 0.09 | 0.01 |
| eccentricity† | 0.039 | 0.040 | 0.025 |
| $\omega$ (radians)† | 0.4 | 1.9 | 2.5 |
| $a/R_*$ | 21.36 | 0.78 | 0.75 |
| $a$ (au) | 0.0327 | 0.0015 | 0.0015 |
| Inclination (deg) | 88.82 | 0.40 | 0.26 |
| Duration (hours) | 1.23 | 0.10 | 0.07 |
| Insolation ($S_\oplus$) | 13.5 | 1.3 | 1.2 |
| **LHS 1678 d** | | | |
| Central transit ($T_c$) (BJD-2457000) | 2000.45806 | 0.00090 | 0.00087 |
| Period (day) | 4.9652229 | 0.0000096 | 0.0000075 |
| Impact parameter† | 0.76 | 0.04 | 0.06 |
| $R_p/R_*$ | 0.0273 | 0.0017 | 0.0018 |
| Radius ($R_\oplus$) | 0.981 | 0.070 | 0.072 |
| K (m s$^{-1}$) | 0.02 | 0.81 | 0.02 |
| eccentricity† | 0.036 | 0.060 | 0.027 |
| $\omega$ (radians])† | 0.4 | 1.9 | 2.3 |
| $a/R_*$ | 26.01 | 0.94 | 0.91 |
| $a$ (au) | 0.0400 | 0.0018 | 0.0017 |
| Inclination (deg) | 88.31 | 0.14 | 0.13 |
| Duration (hr) | 1.00 | 0.11 | 0.08 |
| Insolation ($S_\oplus$) | 9.1 | 0.9 | 0.8 |

In our TRICERATOPS analysis of LHS 1678 d, we used the TESS SPOC PDCSAP light curves at the shortest cadence for all available TESS sectors and the extraction apertures contained in the light-curve headers. These were queried using lightkurve (Lightkurve Collaboration et al. 2018). Since the signal is known to be on target and does not originate from a background star (see Section 2.2), we excluded all nearby stars from false positive consideration. We included the high-resolution imaging contrast curve of the system observed by the Very Large Telescope (VLT) NAOS-CONICA instrument (Lenzen et al. 2003; Rousset et al. 2003) presented by Silverstein et al. (2022) as a constraint on the false positive calculation.

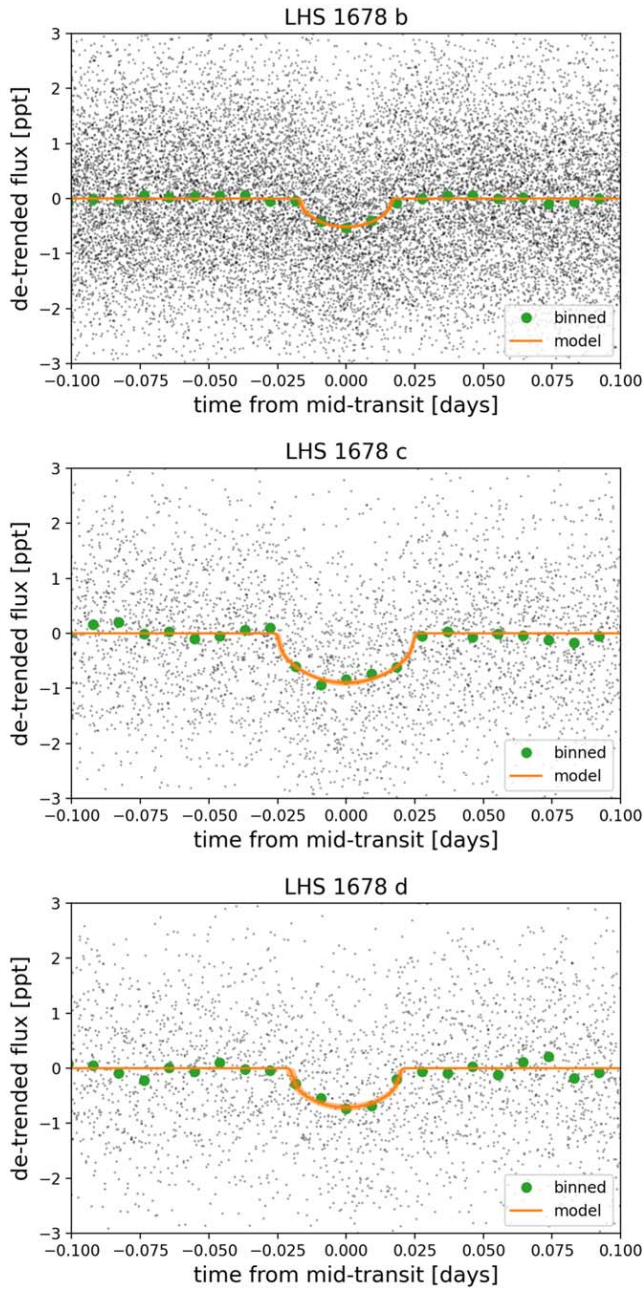

**Figure 1.** TESS photometry from all four sectors folded on orbital period for each planet. Tiny dots are the individual 2 minute cadence measurements, shown also in 13 minute bins as larger green dots. Overlaid as orange curves are the best-fit transit models; shading displays the $1\sigma$ (central 68th percentile) range of models consistent with the data.

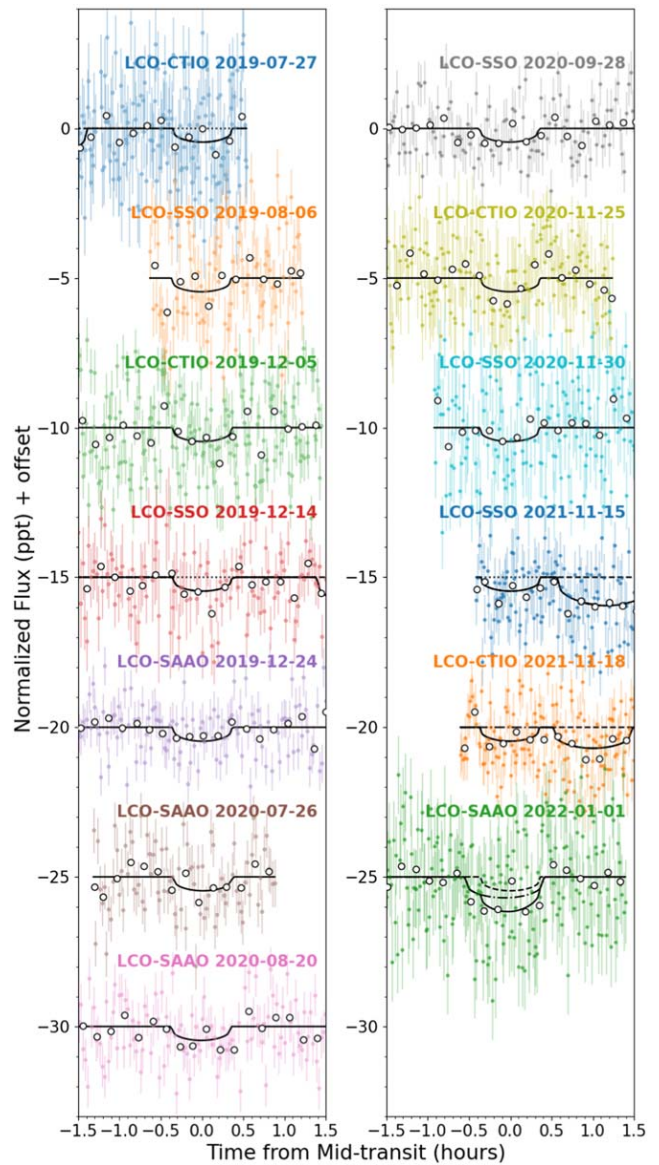

**Figure 2.** Ground-based transit observations of LHS 1678 b in order of their acquisition. Colored points and error bars trace individual, discrete photometric measurements. The white-filled black circles represent the measurements binned to a 10 minute interval. The black lines trace the best-fit model with dashed, dotted, and dotted–dashed lines distinguishing between planets b, c, and d, respectively, in the case of multiple planet transits. The observatory used and date of observation are listed next to each light curve in the same color as the observed data. Light curves taken on 2019 July 27, 2019 December 14, 2021 November 15, 2021 November 18, and 2022 January 1 feature transits of multiple planets, including a partially overlapping transit with LHS 1678 d on 2022 January 1.

Using the above inputs and constraints, TRICERATOPS calculates an FPP of $1.75 \times 10^{-3} \pm 3.39 \times 10^{-4}$, well below the FPP of $1.5 \times 10^{-2}$ needed to consider a target statistically validated (Giacalone et al. 2021). Furthermore, TRICERA-TOPS calculates a nearby FFP (NFFP), or the probability that the source of the signal is a false positive originating from a nearby star. We calculate an NFPP of 0 due to the exclusion of nearby stars from our analysis and determination that the signal is on target. Due to the low FPP and NFPP values from TRICERATOPS, as well as the low FPP from vespa, we consider LHS 1678 d to be a statistically validated planet.

### 4. Prospects for Future Follow-up Observations

Nearby at only 20 pc, the LHS 1678 system is bright and amenable for additional follow-up observations. Here we focus on the potential for further observational characterization of LHS 1678 d and new prospects for the full system. Those specific to LHS 1678 b and c are discussed in Silverstein et al. (2022).

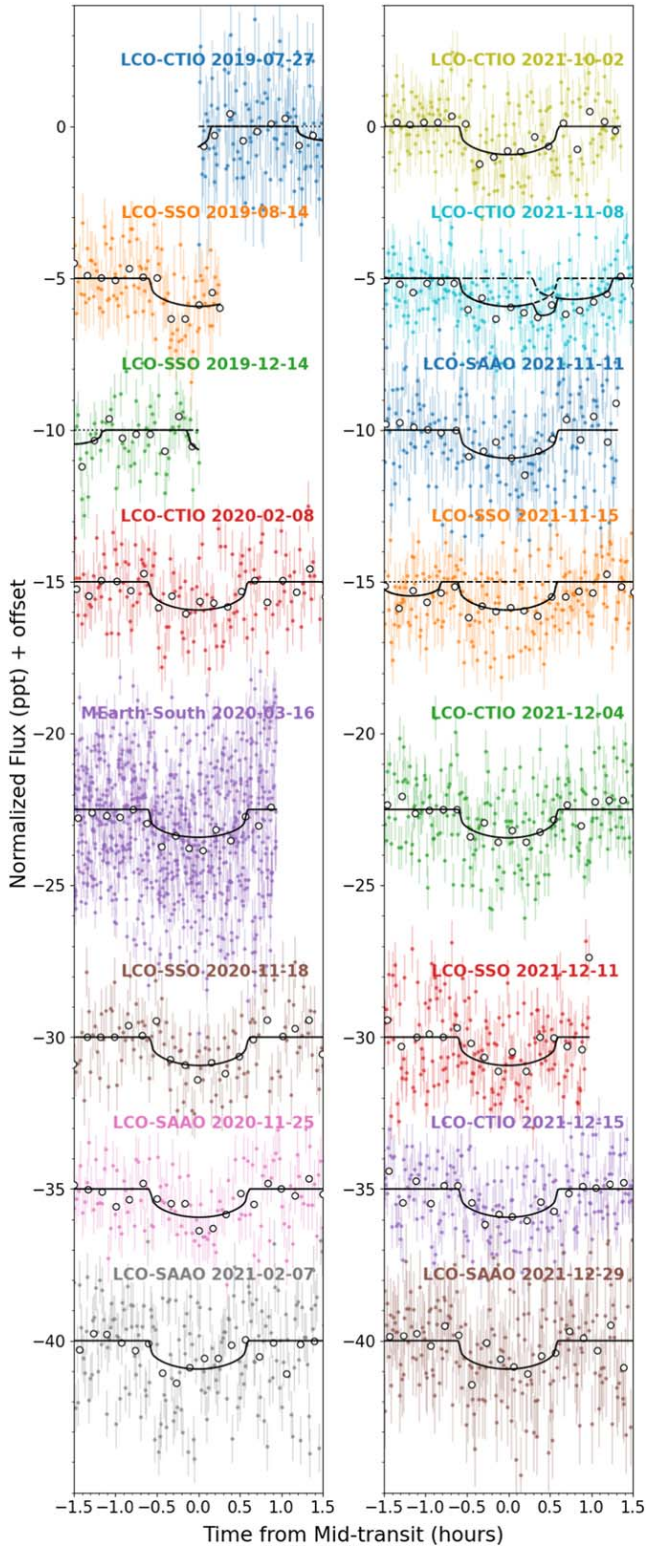

**Figure 3.** Ground-based transit observations of LHS 1678 c from LCO and MEarth, derived and presented in the same way as in Figure 2. Light curves taken on 2019 July 27, 2019 December 14, 2021 November 8, and 2021 November 15 feature multiple and/or simultaneous transits of another planet, including a partially overlapping transit with LHS 1678 d on 2021 November 8.

*Mass.* High-precision radial velocity measurements for LHS 1678 d will be key to determining exoplanet mass and density, which help paint a fuller picture of the system (e.g.,

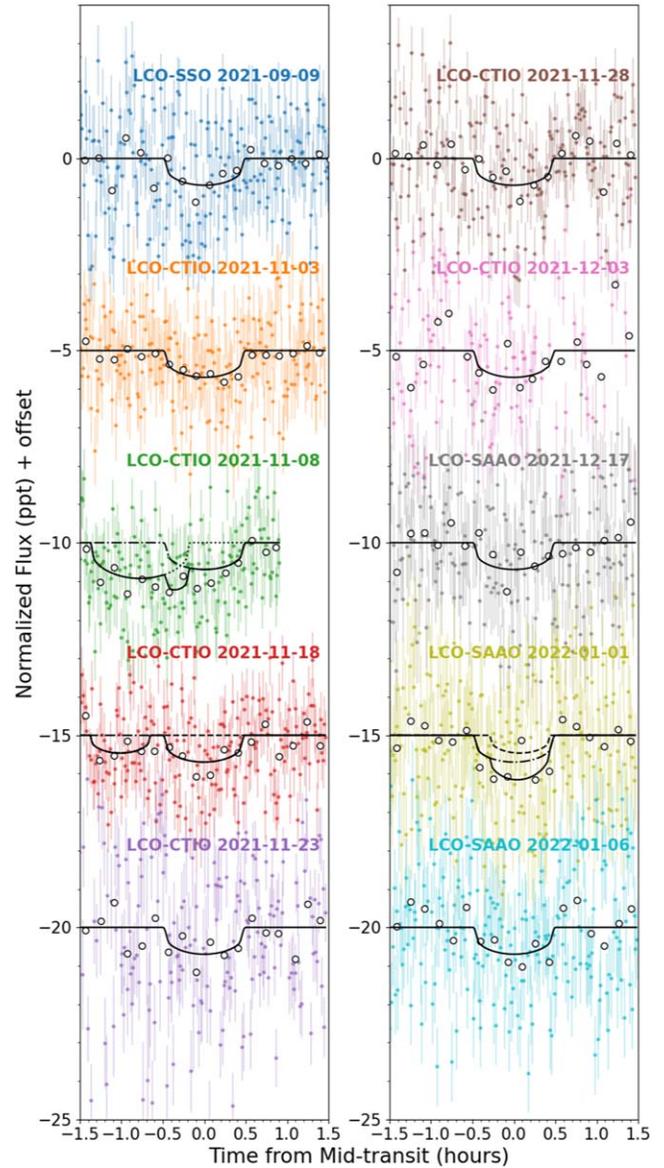

**Figure 4.** Ground-based transit observations of LHS 1678 d from LCO, derived and presented in the same way as in Figures 2 and 3. Light curves on 2021 November 8 and 2022 January 1 demonstrate simultaneous transits, and the light curve on 2021 November 18 shows separate transits of b and d.

Dorn et al. 2015; Luque & Pallé 2022; Goffo et al. 2023). We estimate masses of $0.26^{+0.14}_{-0.10}\,M_\oplus$, $0.81^{+0.55}_{-0.29}\,M_\oplus$, and $0.92^{+0.66}_{-0.34}\,M_\oplus$ for LHS 1678 b, c, and d, respectively, using the `forecaster` procedure (Chen & Kipping 2017). We then predict RV semiamplitudes of 0.4, 0.7, and $0.7\,\mathrm{m\,s^{-1}}$ following Equation (14) in Lovis & Fischer (2010). In our calculation, we assume zero eccentricity for each planet and adopt our reported stellar mass, planetary orbital periods, planetary inclinations, and `forecaster` mass estimates. Our model-based $3\sigma$ upper limits on the radial velocity semiamplitudes for the planets are 3.1, 1.0, and $2.1\,\mathrm{m\,s^{-1}}$, respectively, which are above the anticipated values for these planets (see Section 3.1). However, this result does hint at planet c not being high mass. Given LHS 1678's southerly decl., the VLT equipped with the ESPRESSO instrument is one facility capable of achieving the precision required to detect these planets (Pepe et al. 2010; Suárez Mascareño et al. 2020).

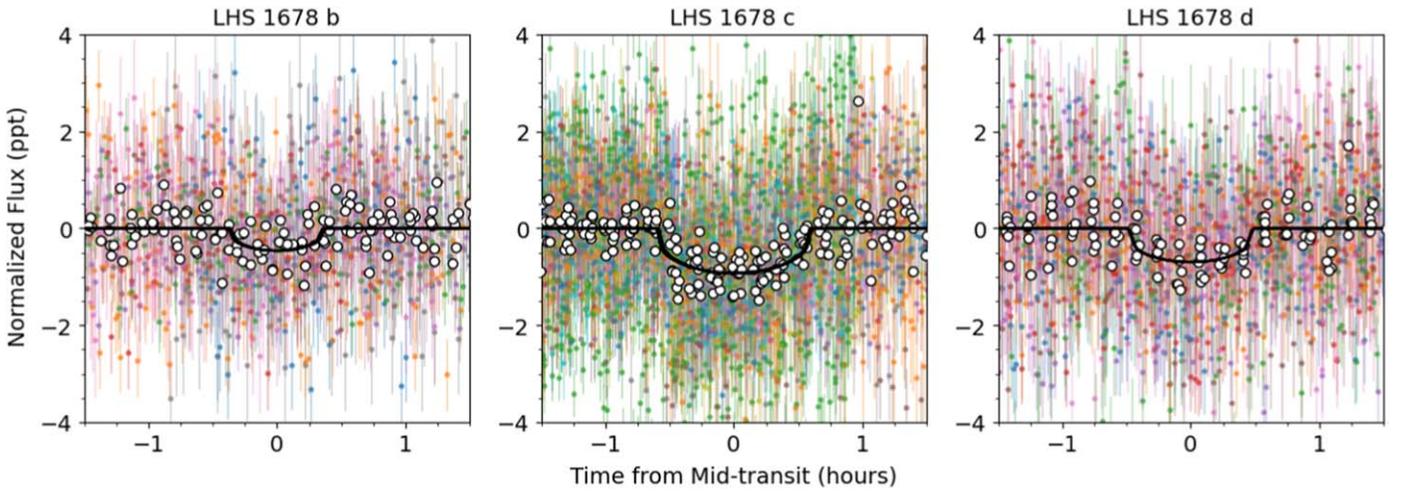

**Figure 5.** To better visualize the successful ground-based recovery of each transiting planet, we stack all ground-based observations, excluding those containing transits of multiple planets. Each transit center is set to 0 hr. Individual transit measurements are plotted using small filled circles with colors matching those shown for each date in Figures 2, 3, and 4. The transits binned to 10 minute intervals are shown as white-filled black circles. The transit models are represented by black lines, with the models for each observation stacked on top of each other.

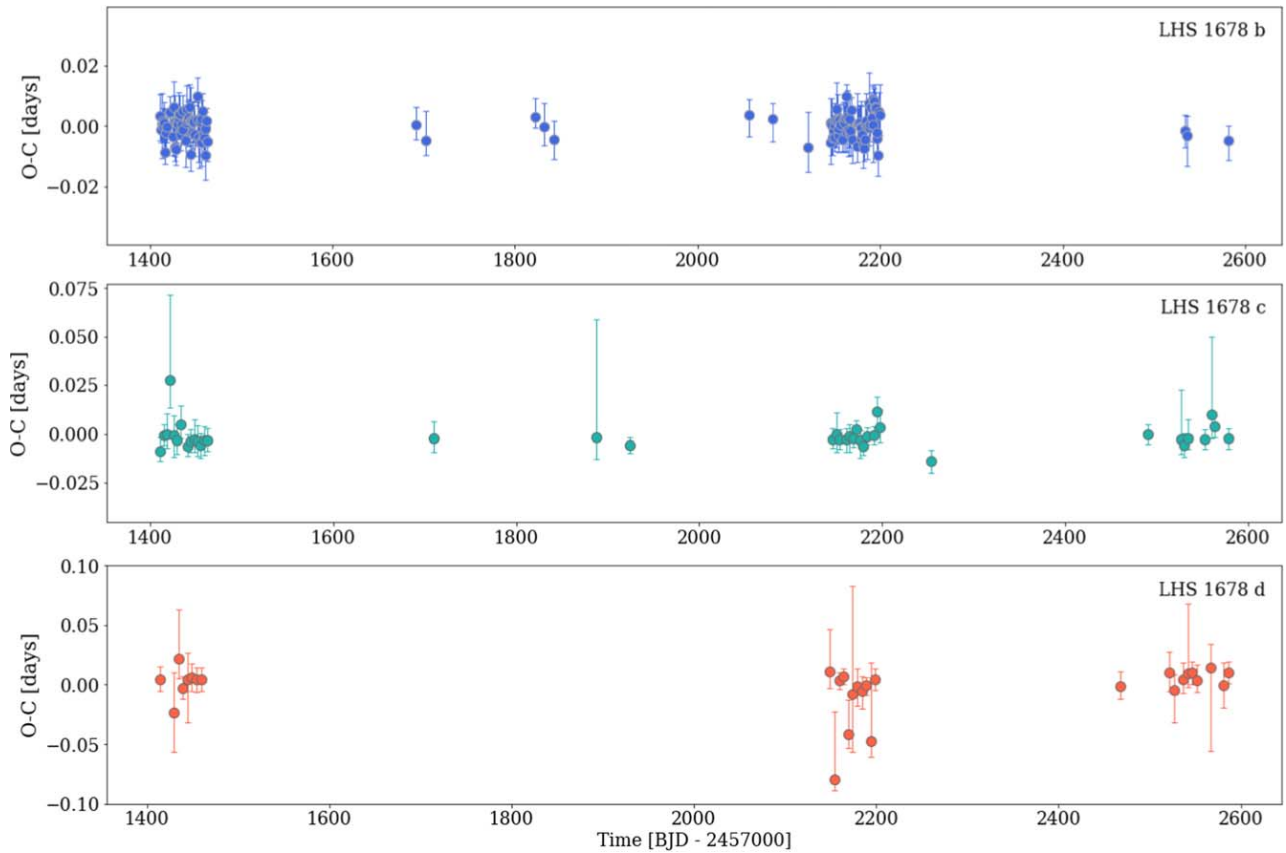

**Figure 6.** Observed minus calculated transit times for LHS 1678b, c, and d reveal no significant signs of TTVs.

*Atmospheric Characterization.* Kempton et al. (2018; henceforth "K18") put forth two metrics to assess the amenability of exoplanets to atmospheric characterization using JWST. In this section, we calculate these metrics using updated parameters for LHS 1678 b and c and for the first time for LHS 1678 d.

Inputting the system parameters reported in this paper, we derive an emission spectroscopy metric (ESM) value of 1.4 for LHS 1678 d, below the minimum favorable value of 7.5 for JWST observations. With their updated properties, we calculate ESM values of 3.9 and 1.9 for LHS 1678 b and c, respectively. These are minimally changed from the Silverstein et al. (2022) values of 3.9 and 2.0. As described by Silverstein et al. (2022), LHS 1678 b is likely more favorable for observations than its ESM suggests. The ESM calculation makes an assumption about the atmospheric properties of the planet, but with its high insolation flux, LHS 1678 b may have little to no atmosphere, which leads to inefficient heat redistribution and more thermal emission from the dayside.

Table 4
Exoplanet Transit Times

| Planet | Transit Time (BJD-2457000) | $+1\sigma$ (day) | $-1\sigma$ (day) | Transit Index | Telescope |
|---|---|---|---|---|---|
| c | 1411.06465716 | 0.00675017 | 0.00520049 | 1 | TESS |
| b | 1411.47688936 | 0.00737215 | 0.00798702 | 1 | TESS |
| b | 1412.33712331 | 0.00447421 | 0.00410170 | 2 | TESS |
| b | 1413.19735726 | 0.01013866 | 0.01130632 | 3 | TESS |
| b | 1414.05759121 | 0.00429192 | 0.00460487 | 4 | TESS |
| d | 1414.56098835 | 0.01055280 | 0.01041387 | 1 | TESS |
| c | 1414.75894158 | 0.00558521 | 0.00524914 | 2 | TESS |
| b | 1414.91782516 | 0.00500581 | 0.00540611 | 5 | TESS |
| b | 1415.77805911 | 0.00690175 | 0.00807824 | 6 | TESS |
| b | 1416.63829306 | 0.00566114 | 0.00414118 | 7 | TESS |
| ⋮ | ⋮ | ⋮ | ⋮ | ⋮ | ⋮ |

**Note.** Note that the index counts the transits, starting from the first predicted transit (1) and increasing with time. We report the corresponding observed transits only, so there are gaps in index.

(This table is available in its entirety in machine-readable form.)

We adopt two masses for our calculation of the transmission spectroscopy metric (TSM). For LHS 1678 d, these are a mass of $0.92\,M_\oplus$ estimated using forecaster and $0.88\,M_\oplus$ as prescribed by K18. With these masses, we derive TSM values of 13.3 and 13.9, respectively. This TSM larger than 10 suggests that LHS 1678 d is favorable for transmission spectroscopy observations using JWST. For LHS 1678 b and c, we employ forecaster masses of 0.26 and $0.81\,M_\oplus$, respectively, and K18 masses of 0.25 and $0.79\,M_\oplus$. The corresponding TSM values are 29.8 and 30.5 for LHS 1678 b and 15.3 and 15.6 for LHS 1678 c. As expected, with the updated planet properties, we only see minimal changes in the TSM values of 30.2 and 15.2 calculated by Silverstein et al. (2022) using K18 masses. The small change is ascribed to small updates in exoplanet radius, which is cubed in the TSM equation and approximately cubed in the K18 mass calculation.

*Comparing Near-twin Planets.* LHS 1678 c and d are near twins in radius and have short-period orbits. Because they are both about the size of the Earth, they are likely rocky in composition. These terrestrial planets also lie in the Venus zone (Kane et al. 2014). Thus, LHS 1678 c and d are valuable targets for comparative exoplanet science. The host star lies in a unique location of the H-R diagram, characterized by an underdensity of low-mass stars. These stars are theorized to mark the boundary between partially and fully convective interiors. As they undergo a core $^3$He burning instability, they are expected to exhibit damped luminosity oscillations with a magnitude of ∼5% over timescales of billions of years (e.g., van Saders & Pinsonneault 2012; Feiden et al. 2021). Due to complex evolution of the convective and radiative portions of the stellar core and interior, our current understanding of magnetic activity evolution may not apply, and early studies are underway to constrain these stars' magnetic properties (Jao et al. 2023). The unique stellar evolution of these H-R diagram gap stars can be more impactful to planets in the Venus zone, where changes in insolation flux and activity may push the climate state of a planet over a threshold (Urquhart et al. 2024, in preparation). The LHS 1678 system provides a rare laboratory to study how two planets with similar properties may have responded differently to unique host star luminosity evolution.

### 5. A Unique System Architecture

The LHS 1678 planets can be examined in the context of the "peas-in-a-pod" trend first identified by Weiss et al. (2018). Key findings were that exoplanets in Kepler (Borucki et al. 2010) multiplanet systems with three or more planets have very similar sizes, uniform semimajor axis spacing, and orbital period ratios that are smaller for smaller planets. The LHS 1678 system contains three terrestrial-sized planets, yet the spacing between the semimajor axes of planets b and c is 2.85 times that between planets c and d. Planet b is also about 30% smaller than planets c and d and has a period ratio of 4.3 with planet c. These properties break the trend for typical Kepler "peas-in-a-pod" systems, but LHS 1678 b is an ultra-short-period planet (USP). It is also possible that there are additional nontransiting planets between planets b and c that may have properties intermediate between the two. There could also be additional planets beyond planet d that are yet to be detected but within reach of current instrumentation. The widely separated brown-dwarf companion identified via astrometric monitoring is not expected to influence the planets unless it is on a highly eccentric orbit. Dynamical modeling by Silverstein et al. (2022) found that the system could be stable with additional, yet-undetected planets.

Weiss et al. (2018) and the majority of studies further exploring the "peas-in-a-pod" phenomenon focus on stars more massive than M dwarfs like LHS 1678. Work is just beginning to understand how this phenomenon varies as a function of key host star properties such as mass (Levine & Laughlin 2023). The LHS 1678 system may play a role in unveiling the differences in how multiplanet systems form and evolve as a function of host star mass, from the Sun-like (0.7–1.6 $M_\odot$) host systems found by Kepler to lower-mass M dwarfs (<0.7 $M_\odot$) such as LHS 1678 at $0.345 \pm 0.014\,M_\odot$ and TRAPPIST-1 at $0.0898 \pm 0.0023\,M_\odot$ (Gillon et al. 2016; Agol et al. 2021).[32]

To investigate further, we examine the current census of confirmed multiplanet systems in the NASA Exoplanet Archive as a function of stellar spectral type, honing in on those with USPs (Figure 7). As of 2023 December 8, the number of multiplanet systems orbiting solar-type stars (here taken to be those with $4000 < T_{\rm eff} \leqslant 6750$ K) is a factor of 7.2 times the number orbiting M-dwarf stars (i.e., those with $T_{\rm eff} \leqslant 4000$ K). Notably, this ratio drops off to only 3.4 when examining only systems with USPs. Cast another way, 9.0% of M-dwarf multiplanet systems contain a USP, compared to only 4.3% for solar-type stars. While there are observational biases that may contribute, these observed statistics are worthy of some consideration. Could there be aspects of multiplanet system formation around low-mass stars that make them more favorable to the production of USPs? A more robust investigation into multiplanet system and USP demographics and formation as a function of spectral type is beyond the scope of this work, but we consider LHS 1678 in the context of recent work on USP systems.

---

[32] Masses are converted from effective temperatures following Pecaut & Mamajek (2013); https://www.pas.rochester.edu/~emamajek/EEM_dwarf_UBVIJHK_colors_Teff.txt.

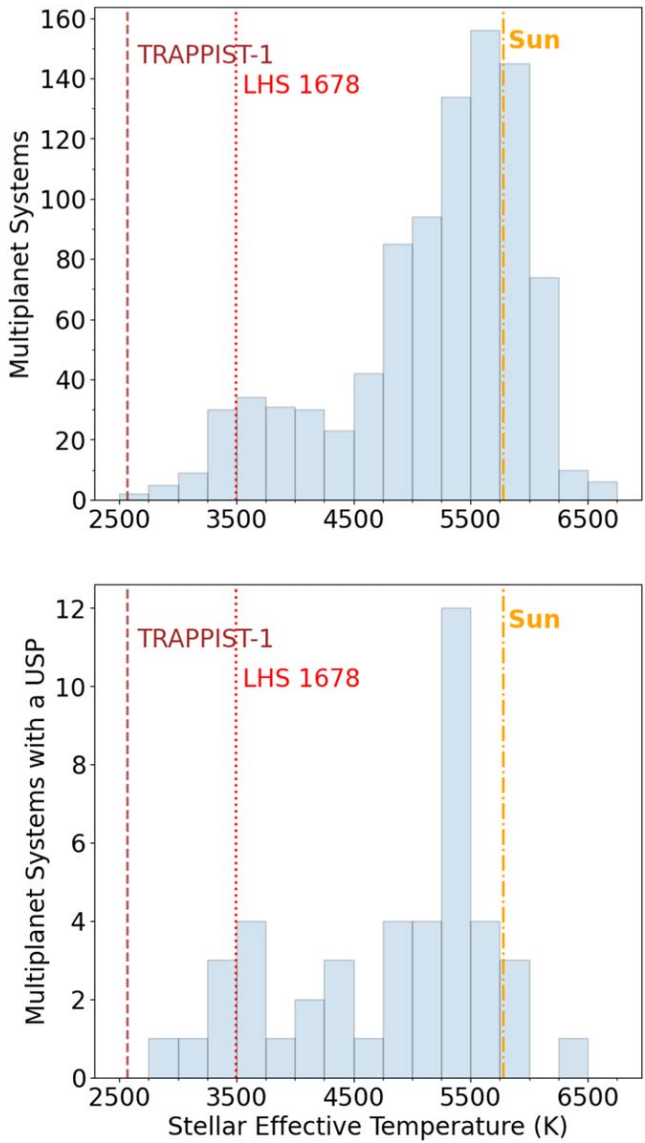

**Figure 7.** Histograms of confirmed multiplanet systems (top panel) and multiplanet systems containing USPs (bottom panel) orbiting stars spanning effective temperatures of approximately 2500–6750 K (NASA Exoplanet Archive; 2023 December 8). This temperature range covers M dwarfs (<4000 K) and solar-type, FGK stars (4000–6750 K). There are far more multiplanet systems around solar-type stars, but this difference narrows for systems that include a USP. One of only 10 multiplanet M-dwarf systems with a USP, LHS 1678 is valuable for investigating planet formation and evolution theories.

Given the above-described census of multiplanet systems, here we briefly consider the ultrashort period of LHS 1678 b and its implications for the formation and evolution of the system. Adams et al. (2021) found that USPs in their sample fell into one of two bins: those with close separations from the next farthest planet (<10 Hill radii) and those with distant separations (>50 Hill radii). The distance between LHS 1678 b and c is about 62 mutual Hill radii, following Equations (3) and (4) of Weiss et al. (2018), who refer to Gladman (1993). Perhaps this separation between the two inner planets is to be expected. However, in the Adams et al. (2021) sample, all the USPs with larger separations were small—approximately Mercury sized, or about half the radius of LHS 1678 b. USPs have also been found to more commonly exhibit higher mutual inclinations with their sibling planets (Dai et al. 2018). Chen et al. (2022) predict that systems containing USPs with lower mutual inclinations likely started coplanar, with possible increases in inclination that decreased over time as the star spun down with age and became less oblate. This suggests that the LHS 1678 planets may have formed in coplanar orbits. LHS 1678 is one of only 10 confirmed, USP-containing, multiplanet systems orbiting an M dwarf as of 2023 December 8. With its resident USP, near-resonant and twin-sized outer planets, and partial "peas-in-a-pod" structure, it serves as a unique system to continue investigating and testing different scenarios of planet formation and evolution in the context of host star properties.

### 6. Conclusion

We have used multiyear TESS monitoring and a large campaign of ground-based transit follow-up observations to validate LHS 1678 d, a third planet orbiting LHS 1678, an M2 dwarf at 20 pc near the H-R diagram gap connected to the M-dwarf fully convective boundary. We have also used these data to update the properties of LHS 1678 b and c, improving their orbit uncertainties and finding that robust mass constraints cannot be made on the LHS 1678 planets with the currently available precision RV data set. We do not further constrain the properties of the brown dwarf companion at this time. LHS 1678 c and d are in near-4:3 orbital resonance, a system property occasionally tied to TTVs. We do not identify TTVs in our 3 yr data set, which includes transits measured by TESS and ground-based facilities. LHS 1678 d joins its sibling LHS 1678 c as a second Earth-sized planet in the Venus zone orbiting this star. These similarities provide a new laboratory for comparative exoplanet analysis in the vicinity of a host star with a unique evolutionary history.

LHS 1678 is one of only 10 confirmed multiplanet M-dwarf systems containing a USP (stars with $T_{\rm eff} \leqslant 4000$ K in the NASA Exoplanet Archive as of 2023 December 8). The USP LHS 1678 b defies the "peas-in-a-pod" trend in the sense that it is considerably smaller and too widely separated from the outer two planets. There may be yet-undetected planets that would stably fill the "pod" though USPs are not always expected to follow the trend. The small relative inclination between LHS 1678 b and its siblings is relatively uncommon and suggests that the system may have formed with coplanar orbits. This combination of system properties and their rarity make LHS 1678 valuable for testing theories of planet formation and evolution. The three LHS 1678 planets are also favorable targets for follow-up mass measurements using high-precision radial velocities and atmospheric characterization using JWST.

## Acknowledgments

This work makes use of observations from the Las Cumbres Observatory global telescope network.

This paper includes data collected with the TESS mission (Ricker et al. 2015), obtained from the MAST data archive (MAST Team 2021a, 2021b, 2021c, 2021d) at the Space Telescope Science Institute (STScI). Funding for the TESS mission is provided by NASAs Science Mission Directorate. STScI is operated by the Association of Universities for Research in Astronomy, Inc., under NASA contract NAS 526555.

We acknowledge the use of TESS High Level Science Products (HLSP) produced by the Quick-Look Pipeline (QLP) at the TESS Science Office at MIT, which are publicly available from the Mikulski Archive for Space Telescopes (MAST). Funding for the TESS mission is provided by NASA's Science Mission directorate. QLP-produced HLSP data can be obtained from the MAST archive (Huang 2020).

We acknowledge the use of public TESS data from pipelines at the TESS Science Office and at the TESS Science Processing Operations Center. Resources supporting this work were provided by the NASA High-End Computing (HEC) Program through the NASA Advanced Supercomputing (NAS) Division at Ames Research Center for the production of the SPOC data products.

This research has made use of the Exoplanet Follow-up Observation Program website, which is operated by the California Institute of Technology, under contract with the National Aeronautics and Space Administration under the Exoplanet Exploration Program.

This work makes use of observations from the LCOGT network. Part of the LCOGT telescope time was granted by NOIRLab through the Mid-Scale Innovations Program (MSIP). MSIP is funded by NSF.

This research made use of `Lightkurve`, a Python package for Kepler and TESS data analysis (Lightkurve Collaboration et al. 2018).

This research made use of `exoplanet` (Foreman-Mackey et al. 2021) and its dependencies (Astropy Collaboration et al. 2013; Kipping 2013; Salvatier et al. 2016; Theano Development Team 2016; Astropy Collaboration et al. 2018; Kumar et al. 2019; Luger et al. 2019; Agol et al. 2020).

This research has made use of the NASA Exoplanet Archive, which is operated by the California Institute of Technology, under contract with the National Aeronautics and Space Administration under the Exoplanet Exploration Program. The Confirmed Planets Table can be accessed in the NASA Exoplanet Archive (NASA Exoplanet Science Institute 2020).

A portion of this work was supported by NASAs Astrophysics Data Analysis Program through grant 20-ADAP20-0016.

The material is based upon work supported by NASA under award number 80GSFC21M0002.

The MEarth Team gratefully acknowledges funding from the David and Lucile Packard Fellowship for Science and Engineering, the National Science Foundation under grants AST-0807690, AST-1109468, AST-1004488, and AST-1616624, the National Aeronautics and Space Administration under grant No. 80NSSC18K0476, and the John Templeton Foundation. The opinions expressed in this publication are those of the authors and do not necessarily reflect the views of the John Templeton Foundation.

This publication was made possible through the support of an LSSTC Catalyst Fellowship to T.D., funded through grant 62192 from the John Templeton Foundation to LSST Corporation. The opinions expressed in this publication are those of the authors and do not necessarily reflect the views of LSSTC or the John Templeton Foundation. This research was supported by the appointment of B.J.H. to the NASA Postdoctoral Program at the NASA Goddard Space Flight Center, administered by Oak Ridge Associated Universities under contract with NASA.

This project was supported in part by an appointment to the NRC Research Associateship Program at the US Naval Research Laboratory, administered by the Fellowships Office of the National Academies of Sciences, Engineering, and Medicine.

*Facilities:* TESS, LCOGT, MEarth, ESO:3.6m, Exoplanet Archive.

*Software:* Python, IDL, `NumPy` (Harris et al. 2020), `Matplotlib` (Hunter 2007), `Astropy` (Astropy Collaboration et al. 2013, 2018, 2022), `Lightkurve` (Lightkurve Collaboration et al. 2018), `TLS` (Hippke & Heller 2019a, 2019b), `AstroImageJ` (Collins et al. 2017), `vespa` (Morton 2015), `exoplanet` (Astropy Collaboration et al. 2013; Kipping 2013; Salvatier et al. 2016; Theano Development Team 2016; Astropy Collaboration et al. 2018; Kumar et al. 2019; Luger et al. 2019; Agol et al. 2020; Foreman-Mackey et al. 2021), `forecaster` (Chen & Kipping 2017), `TESScut` (Brasseur et al. 2019), `Quick-Look Pipeline` (Huang et al. 2020a, 2020b), `TESS-plots` (https://github.com/mkunimoto/TESS-plots).

## ORCID iDs

Michele L. Silverstein ● https://orcid.org/0000-0003-2565-7909
Thomas Barclay ● https://orcid.org/0000-0001-7139-2724
Joshua E. Schlieder ● https://orcid.org/0000-0001-5347-7062
Karen A. Collins ● https://orcid.org/0000-0001-6588-9574
Richard P. Schwarz ● https://orcid.org/0000-0001-8227-1020
Benjamin J. Hord ● https://orcid.org/0000-0003-3904-6754
Jason F. Rowe ● https://orcid.org/0000-0002-5904-1865
Ethan Kruse ● https://orcid.org/0000-0002-0493-1342
Nicola Astudillo-Defru ● https://orcid.org/0000-0002-8462-515X
Xavier Bonfils ● https://orcid.org/0000-0001-9003-8894
Douglas A. Caldwell ● https://orcid.org/0000-0003-1963-9616
David Charbonneau ● https://orcid.org/0000-0002-9003-484X
Ryan Cloutier ● https://orcid.org/0000-0001-5383-9393
Kevin I. Collins ● https://orcid.org/0000-0003-2781-3207
Tansu Daylan ● https://orcid.org/0000-0002-6939-9211
William Fong ● https://orcid.org/0000-0003-0241-2757
Jon M. Jenkins ● https://orcid.org/0000-0002-4715-9460
Michelle Kunimoto ● https://orcid.org/0000-0001-9269-8060
Felipe Murgas ● https://orcid.org/0000-0001-9087-1245
Enric Palle ● https://orcid.org/0000-0003-0987-1593
George R. Ricker ● https://orcid.org/0000-0003-2058-6662
Sara Seager ● https://orcid.org/0000-0002-6892-6948
Avi Shporer ● https://orcid.org/0000-0002-1836-3120
Evan Tey ● https://orcid.org/0000-0002-5308-8603
Roland Vanderspek ● https://orcid.org/0000-0001-6763-6562
Joshua N. Winn ● https://orcid.org/0000-0002-4265-047X

## References

Adams, E. R., Jackson, B., Johnson, S., et al. 2021, PSJ, 2, 152
Agol, E., Dorn, C., Grimm, S. L., et al. 2021, PSJ, 2, 1
Agol, E., Luger, R., & Foreman-Mackey, D. 2020, AJ, 159, 123
Agol, E., Steffen, J., Sari, R., & Clarkson, W. 2005, MNRAS, 359, 567
Astropy Collaboration, Price-Whelan, A. M., Lim, P. L., et al. 2022, ApJ, 935, 167
Astropy Collaboration, Price-Whelan, A. M., Sipőcz, B. M., et al. 2018, AJ, 156, 123
Astropy Collaboration, Robitaille, T. P., Tollerud, E. J., et al. 2013, A&A, 558, A33
Baştürk, Ö., Esmer, E. M., Yalçınkaya, S., et al. 2022, MNRAS, 512, 2062
Berta, Z. K., Irwin, J., Charbonneau, D., Burke, C. J., & Falco, E. E. 2012, AJ, 144, 145
Borucki, W. J., Koch, D., Basri, G., et al. 2010, Sci, 327, 977
Brasseur, C. E., Phillip, C., Fleming, S. W., Mullally, S. E., & White, R. L. 2019, Astrocut: Tools for creating cutouts of TESS images, Astrophysics Source Code Library, ascl:1905.007
Brown, T. M., Baliber, N., Bianco, F. B., et al. 2013, PASP, 125, 1031
Chen, C., Li, G., & Petrovich, C. 2022, ApJ, 930, 58
Chen, J., & Kipping, D. 2017, ApJ, 834, 17
Collins, K. A., Kielkopf, J. F., Stassun, K. G., & Hessman, F. V. 2017, AJ, 153, 77
Dai, F., Masuda, K., & Winn, J. N. 2018, ApJL, 864, L38
Deeg, H. J. 2015, A&A, 578, A17
Dorn, C., Khan, A., Heng, K., et al. 2015, A&A, 577, A83
Feiden, G. A., Skidmore, K., & Jao, W.-C. 2021, ApJ, 907, 53
Ford, E. B., Rowe, J. F., Fabrycky, D. C., et al. 2011, ApJS, 197, 2
Foreman-Mackey, D. 2018, RNAAS, 2, 31
Foreman-Mackey, D., Agol, E., Ambikasaran, S., & Angus, R. 2017, AJ, 154, 220
Foreman-Mackey, D., Luger, R., Agol, E., et al. 2021, JOSS, 6, 3285
Foreman-Mackey, D., Luger, R., Czekala, I., et al. 2020, exoplanet-dev/exoplanet v0.3.2, Zenodo, doi:10.5281/zenodo.1998447
Foreman-Mackey, D., Savel, A., Luger, R., et al. 2021, exoplanet-dev/exoplanet v0.5.0, Zenodo, doi:10.5281/zenodo.1998447
Gardner, J. P., Mather, J. C., Clampin, M., et al. 2006, SSRv, 123, 485
Giacalone, S., & Dressing, C. D. 2020, triceratops: Candidate exoplanet rating tool, Astrophysics Source Code Library, ascl:2002.004
Giacalone, S., Dressing, C. D., Hedges, C., et al. 2022, AJ, 163, 99
Giacalone, S., Dressing, C. D., Jensen, E. L. N., et al. 2021, AJ, 161, 24
Gilbert, E. A., Barclay, T., Schlieder, J. E., et al. 2020, AJ, 160, 116
Gilbert, E. A., Vanderburg, A., Rodriguez, J. E., et al. 2023, ApJL, 944, L35
Gillon, M., Jehin, E., Lederer, S. M., et al. 2016, Natur, 533, 221
Gladman, B. 1993, Icar, 106, 247
Goffo, E., Gandolfi, D., Egger, J. A., et al. 2023, ApJL, 955, L3
Guerrero, N. M., Seager, S., Huang, C. X., et al. 2021, ApJS, 254, 39
Harris, C. R., Millman, K. J., van der Walt, S. J., et al. 2020, Natur, 585, 357
Hippke, M., & Heller, R. 2019a, A&A, 623, A39
Hippke, M., & Heller, R. 2019b, TLS: Transit Least Squares, Astrophysics Source Code Library, ascl:1910.007
Holman, M. J., & Murray, N. W. 2005, Sci, 307, 1288
Huang, C. X., Vanderburg, A., Pál, A., et al. 2020a, RNAAS, 4, 204
Huang, C. X., Vanderburg, A., Pál, A., et al. 2020b, RNAAS, 4, 206
Huang, C. X. 2020, TESS Lightcurves From The MIT Quick-Look Pipeline "(QLP)", STScI/MAST, doi:10.17909/T9-R086-E880
Hunter, J. D. 2007, CSE, 9, 90
Irwin, J. M., Berta-Thompson, Z. K., Charbonneau, D., et al. 2015, in Cambridge Workshop on Cool Stars, Stellar Systems, and the Sun Vol. 18 (Cambridge: Cambridge Univ. Press), 767

Jao, W.-C., Henry, T. J., Gies, D. R., & Hambly, N. C. 2018, ApJL, 861, L11
Jao, W.-C., Henry, T. J., White, R. J., et al. 2023, AJ, 166, 63
Jao, W.-C., Henry, T. J., Winters, J. G., et al. 2017, AJ, 154, 191
Jenkins, J. M., Chandrasekaran, H., McCauliff, S. D., et al. 2010, Proc. SPIE, 7740, 77400D
Jenkins, J. M., Twicken, J. D., McCauliff, S., et al. 2016, Proc. SPIE, 9913, 99133E
Kane, S. R., Kopparapu, R. K., & Domagal-Goldman, S. D. 2014, ApJL, 794, L5
Kempton, E. M. R., Bean, J. L., Louie, D. R., et al. 2018, PASP, 130, 114401
Kipping, D. M. 2013, MNRAS, 435, 2152
Kovács, G., Zucker, S., & Mazeh, T. 2002, A&A, 391, 369
Kumar, R., Carroll, C., Hartikainen, A., & Martin, O. A. 2019, JOSS, 4, 1143
Lenzen, R., Hartung, M., Brandner, W., et al. 2003, Proc. SPIE, 4841, 944
Levine, W., & Laughlin, G. 2023, BAAS, 55, 344.04
Lightkurve Collaboration, Cardoso, J. V. D. M., Hedges, C., et al. 2018, Lightkurve: Kepler and TESS time series analysis in Python, Astrophysics Source Code Library, ascl:1812.013
Lovis, C., & Fischer, D. 2010, in Exoplanets, ed. S. Seager (Tucson, AZ: Univ. Arizona Press), 27
Luger, R., Agol, E., Foreman-Mackey, D., et al. 2019, AJ, 157, 64
Luque, R., & Pallé, E. 2022, Sci, 377, 1211
MAST Team 2021a, TESS Light Curves—All Sectors, STScI/MAST, doi:10.17909/T9-NMC8-F686
MAST Team 2021b, TESS Target Pixel Files—All Sectors, STScI/MAST, doi:10.17909/T9-YK4W-ZC73
MAST Team 2021c, TESS Data Validation Files—All Single-Sectors, STScI/MAST, doi:10.17909/T9-2TC5-A751
MAST Team 2021d, TESS Data Validation Files—All Multi-Sectors, STScI/MAST, doi:10.17909/T9-YJJ5-4T42
Mayor, M., Pepe, F., Queloz, D., et al. 2003, Msngr, 114, 20
McCully, C., Volgenau, N. H., Harbeck, D.-R., et al. 2018, Proc. SPIE, 10707, 107070K
Ment, K., & Charbonneau, D. 2023, AJ, 165, 265
Morton, T. D. 2015, VESPA: False positive probabilities calculator, Astrophysics Source Code Library, ascl:1503.011
Morton, T. D., Giacalone, S., & Bryson, S. 2023, RNAAS, 7, 107
Muirhead, P. S., Dressing, C. D., Mann, A. W., et al. 2018, AJ, 155, 180
NASA Exoplanet Science Institute 2020, Planetary Systems Table, IPAC
Nutzman, P., & Charbonneau, D. 2008, PASP, 120, 317
Pecaut, M. J., & Mamajek, E. E. 2013, ApJS, 208, 9
Pepe, F., Mayor, M., Rupprecht, G., et al. 2002, Msngr, 110, 9
Pepe, F. A., Cristiani, S., Rebolo Lopez, R., et al. 2010, Proc. SPIE, 7735, 77350F
Ricker, G. R., Winn, J. N., Vanderspek, R., et al. 2015, JATIS, 1, 014003
Rodríguez Martínez, R., Martin, D. V., Gaudi, B. S., et al. 2023, AJ, 166, 137
Rousset, G., Lacombe, F., Puget, P., et al. 2003, Proc. SPIE, 4839, 140
Salvatier, J., Wiecki, T. V., & Fonnesbeck, C. 2016, PeerJ Comp. Sci., 2, e55
Silverstein, M. L., Schlieder, J. E., Barclay, T., et al. 2022, AJ, 163, 151
Smith, J. C., Stumpe, M. C., Van Cleve, J. E., et al. 2012, PASP, 124, 1000
Stassun, K. G., Oelkers, R. J., Pepper, J., et al. 2018, AJ, 156, 102
Stumpe, M. C., Smith, J. C., Catanzarite, J. H., et al. 2014, PASP, 126, 100
Stumpe, M. C., Smith, J. C., Van Cleve, J. E., et al. 2012, PASP, 124, 985
Suárez Mascareño, A., Faria, J. P., Figueira, P., et al. 2020, A&A, 639, A77
Theano Development Team 2016, arXiv:1605.02688
van Saders, J. L., & Pinsonneault, M. H. 2012, ApJ, 751, 98
Vanderburg, A., Latham, D. W., Buchhave, L. A., et al. 2016, ApJS, 222, 14
Weiss, L. M., Marcy, G. W., Petigura, E. A., et al. 2018, AJ, 155, 48